\documentclass[aps,pra,onecolumn,amssymb,amsfonts,amsmath,showpacs,preprint,a4paper,superscriptaddress]{revtex4}

\usepackage{float}
\usepackage{color}
\usepackage[final]{graphicx}

\newcommand{\be}{\begin{equation}}
\newcommand{\ee}{\end{equation}}
\newcommand{\bea}{\begin{eqnarray}}
\newcommand{\eea}{\end{eqnarray}}

\begin{document}
\title{Doping dependence and electron-boson coupling in the ultrafast relaxation of hot electron populations in Ba(Fe$_{1-x}$Co$_x$)$_2$As$_2$}

\author{I. Avigo}
\affiliation {Fakult\"{a}t f\"{u}r Physik, Universit{\"a}t Duisburg-Essen, Lotharstr. 1, 47057 Duisburg, Germany}

\author{S. Thirupathaiah}
\altaffiliation[current address: ]{Solid State and Structural Chemistry Unit, Indian Institute of Science, Bangalore, Karnataka, 560012, India}
\affiliation{Fakult\"{a}t f\"{u}r Physik, Universit{\"a}t Duisburg-Essen, Lotharstr. 1, 47057 Duisburg, Germany}



\author{M. Ligges}
\affiliation {Fakult\"{a}t f\"{u}r Physik, Universit{\"a}t Duisburg-Essen, Lotharstr. 1, 47057 Duisburg, Germany}

\author{T. Wolf}
\affiliation{Karlsruhe Institute of Technology, Institut f\"{u}r Festk\"{o}rperphysik, 76021 Karlsruhe, Germany}

\author{J. Fink}
\affiliation{Leibniz-Institute for Solid State and Materials Research Dresden, P.O.Box 270116, 01171 Dresden, Germany}

\author{U. Bovensiepen}
\email{uwe.bovensiepen@uni-due.de}
\homepage {www.uni-due.de/agbovensiepen}
\affiliation {Fakult\"{a}t f\"{u}r Physik, Universit{\"a}t Duisburg-Essen, Lotharstr. 1, 47057 Duisburg, Germany}

\date{\today}

\begin{abstract}

\footnotesize{Using femtosecond time- and angle-resolved photoemission spectroscopy we investigate the effect of electron doping on the electron dynamics in Ba(Fe$_{1-x}$Co$_x$)$_2$As$_2$} in a range of $0\leq x<0.15$ at temperatures slightly above the N\'eel temperature. By analyzing the time-dependent photoemission intensity of the pump laser excited population as a function of energy, we found that the relaxation times at $0<E-E_\text{F}<0.2\,\mathrm{eV}$ are doping dependent and about 100 fs shorter at optimal doping than for overdoped and parent compounds. Analysis of the relaxation rates also reveals the presence of a pump fluence dependent step in the relaxation time at $E-E_\text{F}=200\,\mathrm{meV}$ which we explain by coupling of the excited electronic system to a boson of this energy. We compare our results with static ARPES and transport measurements and find disagreement and agreement concerning the doping-dependence, respectively. We discuss the effect of the electron-boson coupling on the energy-dependent relaxation and assign the origin of the boson to a magnetic excitation.
\end{abstract}
\maketitle

\section{\label{Intro}Introduction}

 The discovery of Fe-based high-$T_c$ superconductivity is one of the breakthroughs in the last two decades of condensed matter research~\cite{Kamihara2008} because in addition to the cuprates  scientists identified a second class of compounds displaying superconductivity with high $T_c$ enhancing the interest and hope in explaining this phenomenon. However, the richness of the physics contained in these materials does not make things smoother and continuously triggers new challenges in the research~\cite{Chubkov(2015)}.
 Understanding the microscopic origins of electronic phases in high-$T_c$ superconductors (HTSC) is important for elucidating the mechanism of superconductivity.\\
 At high temperatures the parent compounds are bad paramagnetic metals with a Fermi surface consisting of three hole pockets at the Brillouin zone center ($\Gamma$) and two electron pockets at the zone corner ($X/M$)~\cite{Fink(2009), Singh(2008)}. The strong nesting between the pockets leads, at temperatures below the N\'eel temperature $T_N$, to a transition into an antiferromagnetic (AFM) state, where the Fe spins assume a spin density wave ordering accompanied by a simultaneous or slightly precedent structural tetragonal-to-orthorhombic transition~\cite{DLC(2008), Huang(2008)}. A consequence of the magnetic ordering is the backfolding of the electron bands to $\Gamma$ where they hybridize strongly with the non-folded states, leading to the opening of energy gaps ~\cite{deJong(2010)}. Before the AFM order is completly established, nematic order sets in, characterized by orbital anisotropy~\cite{Yi(2011)} and quantum fluctuations that persist above the structural transition ~\cite{Lu(2014), Harriger(2011)}. Superconductivity arises by tuning a control parameter, as doping or pressure, that gradually reduces the AFM order until it is suppressed.
 A mainstream view is that the strong antiferromagnetic fluctuations at this point, called a quantum critical point (QCP), could be a good candidate for the pairing mechanism that leads to superconductivity and that these fluctuations would also account for the normal state non-Fermi-liquid (NFL) behavior. A QCP has been theoretically predicted~\cite{Dai(2009), Abraham(2011)}, measurements of the London penetration depth~\cite{Hashimoto(2012)} support a quantum critical point of view and experimental efforts have observed anomalous normal state electronic properties in the resistivity ~\cite{Chu(2010),Analytis(2014)}, optical conductivity~\cite{Nakajima(2011)}, thermal ~\cite{Meingast(2012)} and NMR~\cite{Nakai(2013)} studies that would explain the NFL-behavior of the metallic state and therefore support the quantum critical scenario.
 On the other hand, there are other models explaining the normal state properties of HTSC~\cite{Kastrinakis(1999)} and our own recent ARPES study~\cite{Fink(2016)} showed no evidence of quantum critical features but rather of a BCS-Bose-Einstein crossover state in the superconductive phase.

 In this frame, further study is required and a valuable contribution could come from other experimental techniques like, for instance, doping-dependent time-resolved (tr) studies, which are lacking in this field. Tr-optical reflectivity~\cite{Mansart(2009)} and trARPES~\cite{Rettig(2012)} had been important for the study of the phonons in the system under non-equilibrium conditions after fs laser excitation and the determination of the electron-phonon coupling strength~\cite{Mansart(2010),Avigo(2012),Rettig(2013)}, pointing out that a pairing mediated by phonons is unlikely and the attention had to be turned to other "non-conventional" mechanisms.
 In particular trARPES is a powerful and versatile tool combining the direct access to the electronic structure in momentum space with the non-equilibrium regime excited by femtosecond laser pulses and probing the ultrafast relaxation dynamics~\cite{Bov(2012)}.
 In the work presented here, we report on a systematic trARPES study on parent ($T_N=140\,\mathrm{K}$) and Co-substituted, i.e. electron doped BaFe$_2$As$_2$ with optimal ($x=0.06$, $T_c=23\,\mathrm{K}$) and overdoped ($x=0.14$) concentration, to investigate the influence of doping on the dynamics of the system. By analyzing the hot electron population decay as a function of energy above $E_{\text{F}}$ we find a trend of the relaxation for different doping in agreement with transport measurements~\cite{Katayama(2009), Rullier-Albenque(2009)} and a blocking of the relaxation at energies below $200\,\mathrm{meV}$. The latter indicates coupling to a boson which is most probably of magnetic origin. We discuss our results in the context of the latest developed scenarios.


\section{\label{exp}Experimental details}

Samples are grown by a self-flux method~\cite{Hardy(2010)} and cleaved in ultrahigh vacuum ($p<10^{-10}\,\mathrm{mbar}$) at a temperature equal or lower than $80\,\mathrm{K}$. Pump-probe measurements were performed after having increased the temperature to $160\,\mathrm{K}$. A commercial regenerative Ti:Sapphire amplifier (Coherent RegA 9040) generates infrared laser pulses with $800\,\mathrm{nm}$ central wavelength, an energy of $1.5\,\mathrm{eV}$ with a repetition rate of $250\,\mathrm{kHz}$ and $40\,\mathrm{fs}$ pulse duration. The $\sim 6\,\mathrm{\mu J}$ pulses are split and 50 percent is used as pump, while the other half of the beam is frequency doubled twice and compressed using prisms to obtain the ultraviolet (UV) fourth harmonic probe of $6\,\mathrm{eV}$ and $80\,\mathrm{fs}$ pulse duration. The time-delay between the pump and the probe beams is achieved through a mechanical delay stage. After the pump pulse has excited the sample, photoelectrons with a certain energy and momentum are generated by the UV pulses and detected by a time-of-flight (TOF) spectrometer with an acceptance angle of $\pm3.8^o$, providing direct access to the dynamics of excited electrons in the vicinity of the Fermi level $E_{\text{F}}$. A schematic of the trARPES experiment is sketched in Fig. \ref{fig:fig1} (a). The overall time resolution was $<90\,\mathrm{fs}$ and the overall energy resolution, determined by the TOF spectrometer and laser bandwidth, was about $50\,\mathrm{meV}$. For details about the experimental setup see Refs.~\cite{Syed(2015),Ligges(2014)}. Photoelectron detection was carried out in normal emission in the vicinity of the $\Gamma$ point at $T=160\,\mathrm{K}$, so that all the compounds have been initially, before optical excitation, in the paramagnetic phase and the only tuned parameter was the electron doping concentration.

\section{\label{exp_res}Experimental results}


\begin{figure}
\includegraphics[width=0.5\textwidth]{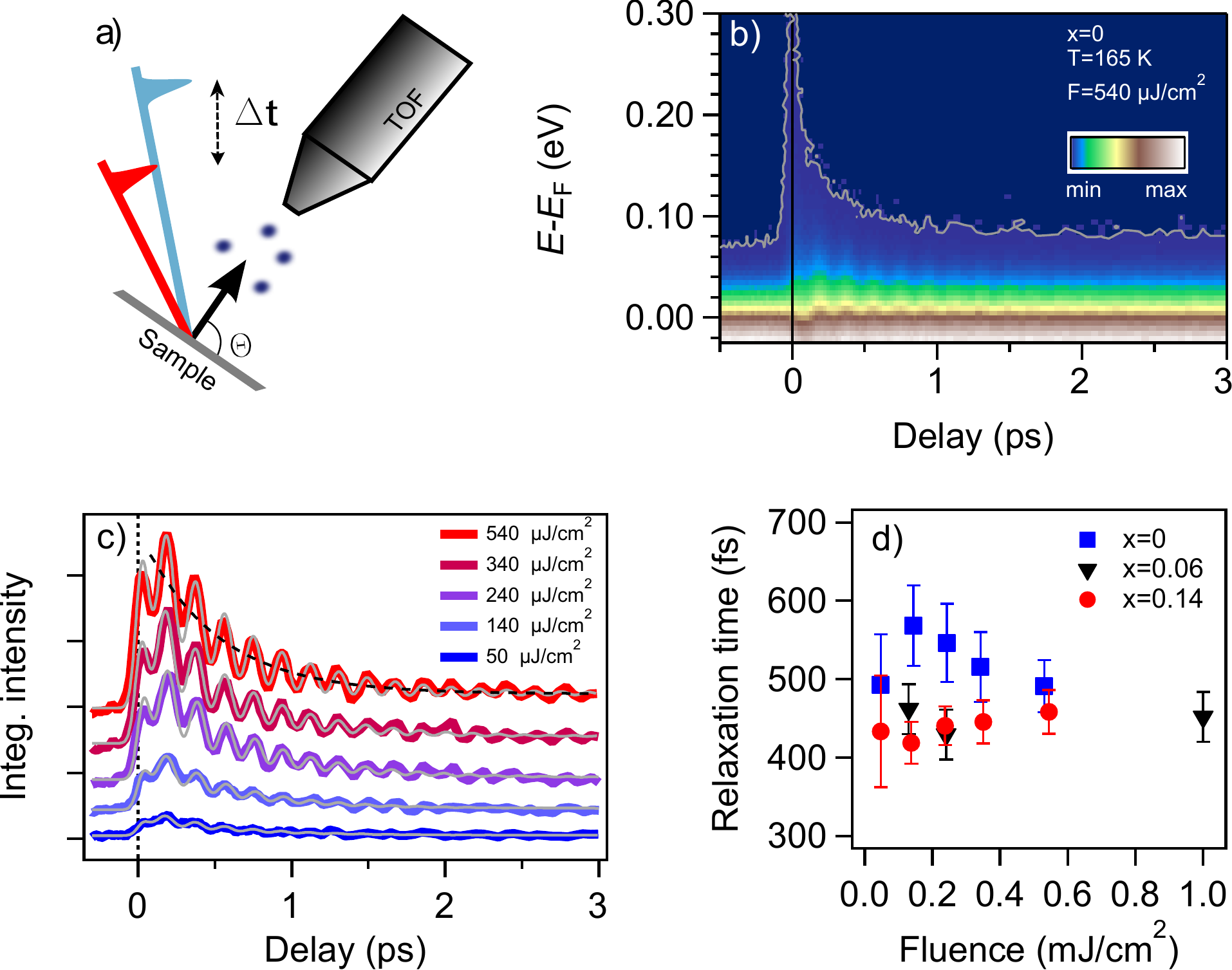}%
\caption{\label{fig:fig1}(color). \footnotesize{a) Schematic of the time-resolved photoemission experiment: after the IR pump pulse has brought the sample out of equilibrium, a time delayed UV probe pulse generates photoelectrons, detected by a time of flight spectrometer. We analyze the dynamics of the excited electrons in the vicinity of the Fermi level.  b) trARPES intensity of BaFe$_2$As$_2$ around the Fermi level as a function of $E-E_{\text{F}}$ and pump-probe delay. The grey contour has been added to further evidence the pump-induced changes. c) Integrated tr-ARPES intensity over an energy range of $\sim1\,\mathrm{eV}$ above $E_{\text{F}}$ for different fluences as indicated; solid grey lines are fits to the data and the dashed black line shows the exponential decay component of the fit. d) Relaxation time constants extracted from the fits in c) for the investigated doping concentrations are plotted as a function of incident pump-fluence.}}
\end{figure}

Exemplary trARPES spectra are shown in Fig. \ref{fig:fig1} (b) for the parent compound at an incident laser pump fluence $F=540\,\mathrm{\mu J/cm^2}$. Upon photoexcitation we observe an increase of spectral weight above $E_{\text{F}}$ up to $1\,\mathrm{eV}$ generated by electrons excited from outside the hole-pockets and subsequent relaxation dominated by cooling of hot electrons to the lattice, mediated by electron-phonon coupling.
Coherent oscillations superimposed to the incoherent exponential decay are caused by the dominant excitation of the $A_{1g}$ phonon mode~\cite{Avigo(2012), Yang(2014)}. In this work we focus on the analysis of the relaxation mechanism of the excited electrons as function of doping concentration and start here by investigating pump fluences of $50-540\,\mathrm{\mu J/cm^2}$.
To analyze the relaxation of the excited electrons we integrated the photoelectron intensity in an energy range from $E_{\text{F}}$ to about $1.5\,\mathrm{eV}$. We describe the obtained time-dependent data with a single exponential decay plus a damped sine function (to account for the coherent oscillations) convoluted with the temporal pump-probe envelope,  see Fig.\ref{fig:fig1} (c).
The resulting exponential decay fits are in the sub-picosecond timescale, in agreement with ~\cite{Rettig(2012), Mansart(2010)} and are summarized in Fig. \ref{fig:fig1} (d) for all samples and for different incident pump fluences. At low fluences, up to $\sim400\,\mathrm{\mu J/cm^2}$, the parent compound relaxes clearly slower than the Co-substituted compounds, showing also a dependence on fluence where, except for the first data point, we see a decrease from $570\,\mathrm{fs}$ to $490\,\mathrm{fs}$ for more than a tripling of $F$. Such an $F$-dependence is not observed in this analysis in the case of the optimally doped (OP) and overdoped (OD) compounds, where the relaxation times remain constant but even up to $200\,\mathrm{fs}$ shorter at $140\,\mathrm{\mu J/cm^2}$. At higher fluences different relaxation times for different doping merge to joint relaxation times a bit below $500\,\mathrm{fs}$.

\begin{figure}
\includegraphics[width=0.5\textwidth]{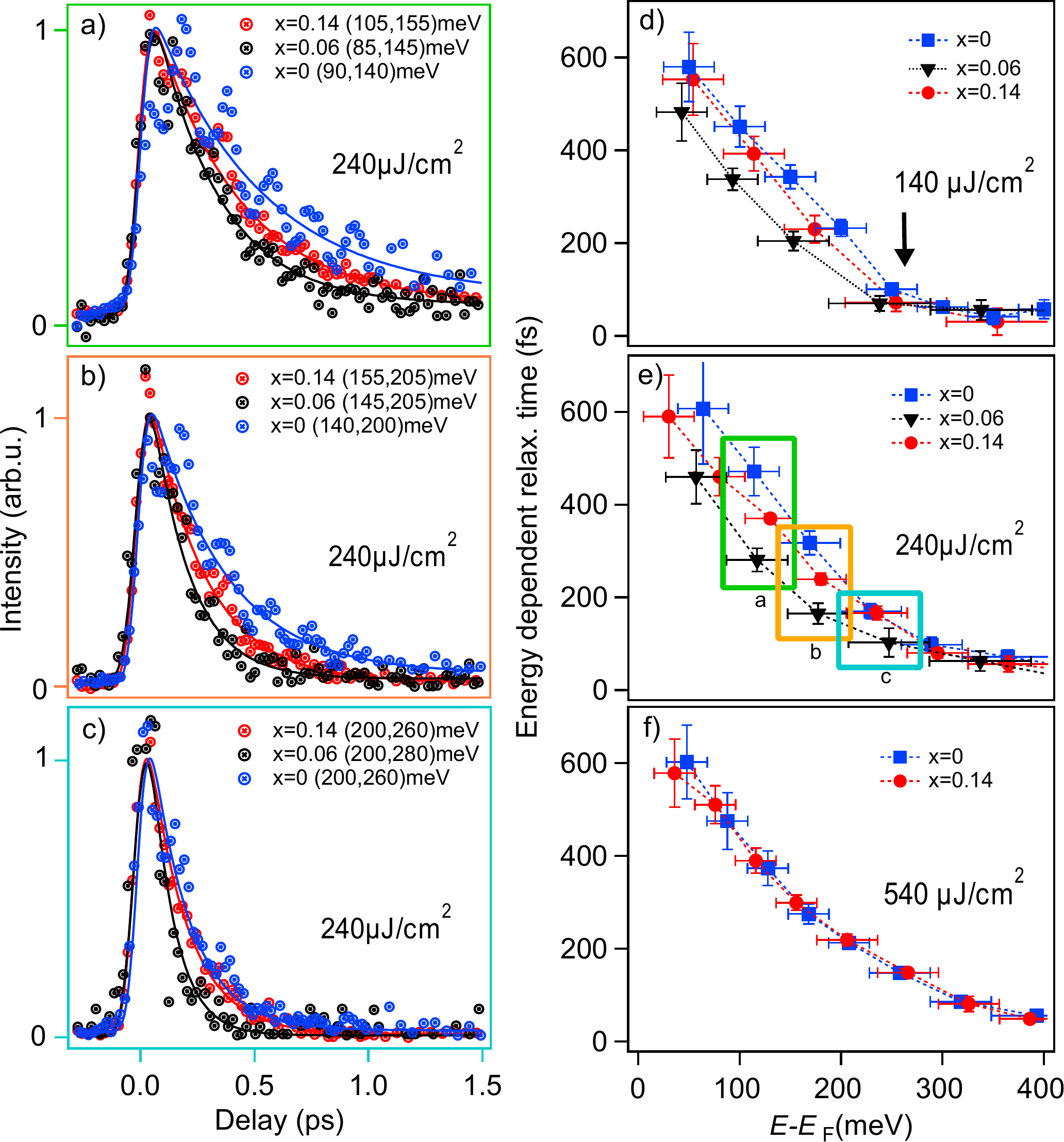}
\caption{\label{fig:fig2}\footnotesize{a)-c) normalized time-dependent intensity extracted in three consecutive energy interval starting from $\sim100\,\mathrm{meV}$ above $E_{\text{F}}$ to $\sim250\,\mathrm{meV}$ at $F=240\,\mathrm{\mu J/cm^2}$. For each energy interval, all three doping concentration are shown. Solid lines are exponential fits to the data. d)-f) Relaxation times extracted from the fit for all curves at three different fluences as a function of binding energy. Error bars on the energy axis mark the energy intervals of integration, the colored boxes, as well as the letters, in Fig.\ref{fig:fig2} (e) compare the results of the fits in comparable energy windows for the three dopings to the corresponding experimental data and fit in Fig. \ref{fig:fig2} (a-c). The integration windows for the curve slightly differ from each others in order to match statistics. The colored boxes in e) represent the corresponding results of the fits reported in a)-c) as indicated by the frame colors and letters.}}
\end{figure}

At this point we emphasize that the previous analysis accounts for the decay of the energy integrated hot electron population. Contribution of electron-electron ($e-e$) scattering at higher energies and the respective secondary electrons closer to $E_\text{F}$ contribute in a far from trivial manner in addition to electron-phonon scattering which mediates close to $E_\text{F}$ energy transfer out of the electron system into phonons~\cite{Rettig(2013), Perfetti(2007), Bov(2007), Ligges(2009)}. The relaxation times reported in Fig.\ref{fig:fig1}(d) are considered as an effective characteristic quantity that cannot be compared with energy and momentum-dependent single particle scattering rates determined from linewidth analysis in static ARPES. The reason is that ARPES line widths are determined by i) elastic and inelastic scattering processes of ii) individual quasiparticle excitations, while trARPES analyzes the population decay to which secondary electrons contribute in a non-trivial manner here after energy integrating the transient population~\cite{Petek(1997), Yang(2015)}. We will come back to this point below and remark that nevertheless these effective relaxation times (Fig \ref{fig:fig1}d) already exhibit a clear doping dependence.
Further insight can be gained, as we show in the following, from an energy resolved analysis, though the difference between static and time-resolved ARPES will persist. We analyze time-dependent photoemission intensities in different energy windows starting from $20\,\mathrm{meV}$ to about $1\,\mathrm{eV}$ above $E_{\text{F}}$, to obtain energy-dependent relaxation times ($\tau$) using a single exponential decay fit function $I=A \exp(t/\tau)+B$ convoluted with a Gaussian function, where $A$ is the excitation amplitude, $\tau$ the relaxation time and $B$ is a constant accounting for lattice heating. A detailed inspection of Fig. \ref{fig:fig1}b provides support for this analysis. At $E-E_\text{F}=0.1-0.3\,\mathrm{eV}$ an energy dependent relaxation is observed, while below $0.1\,\mathrm{eV}$, where lattice heating will take an effect on the electron distribution, the intensity barely decays up to 3 ps. Some exemplary curves for the fluence of $240\,\mathrm{\mu J/cm^2}$ are shown in Fig. \ref{fig:fig2}(a)-(c) comparing dynamics of samples with different doping concentration in similar energy windows. The results of the complete analysis on all compounds are shown in Fig. \ref{fig:fig2}(d)-(f).


We observe a doping dependence until $E-E_{\text{F}}=250\,\mathrm{meV}$. As for the energy integrated intensities, the parent compound relaxes on a longer timescale than the Co-substituted compounds but it becomes now clear that the $x=0.06$ (optimally doped) relaxes faster than the $x=0.14$ (overdoped) compound. If we compare, for example, $\tau$ at $E-E_\text{F}=0.1\,\mathrm{eV}$ we get $280\,\mathrm{fs}$ for the $x=0.06$, while we obtain for $x=0.14$ and parent $370\,\mathrm{fs}$ and $470\,\mathrm{fs}$, respectively. These differences in $\tau$ in the first $250\,\mathrm{meV}$ are clearly visible at $140\,\mathrm{\mu J/cm^2}$ and $240\,\mathrm{\mu J/cm^2}$ but they vanish at the highest $F$ of $540\,\mathrm{\mu J/cm^2}$. \\


\begin{figure}
\includegraphics[width=0.5\textwidth]{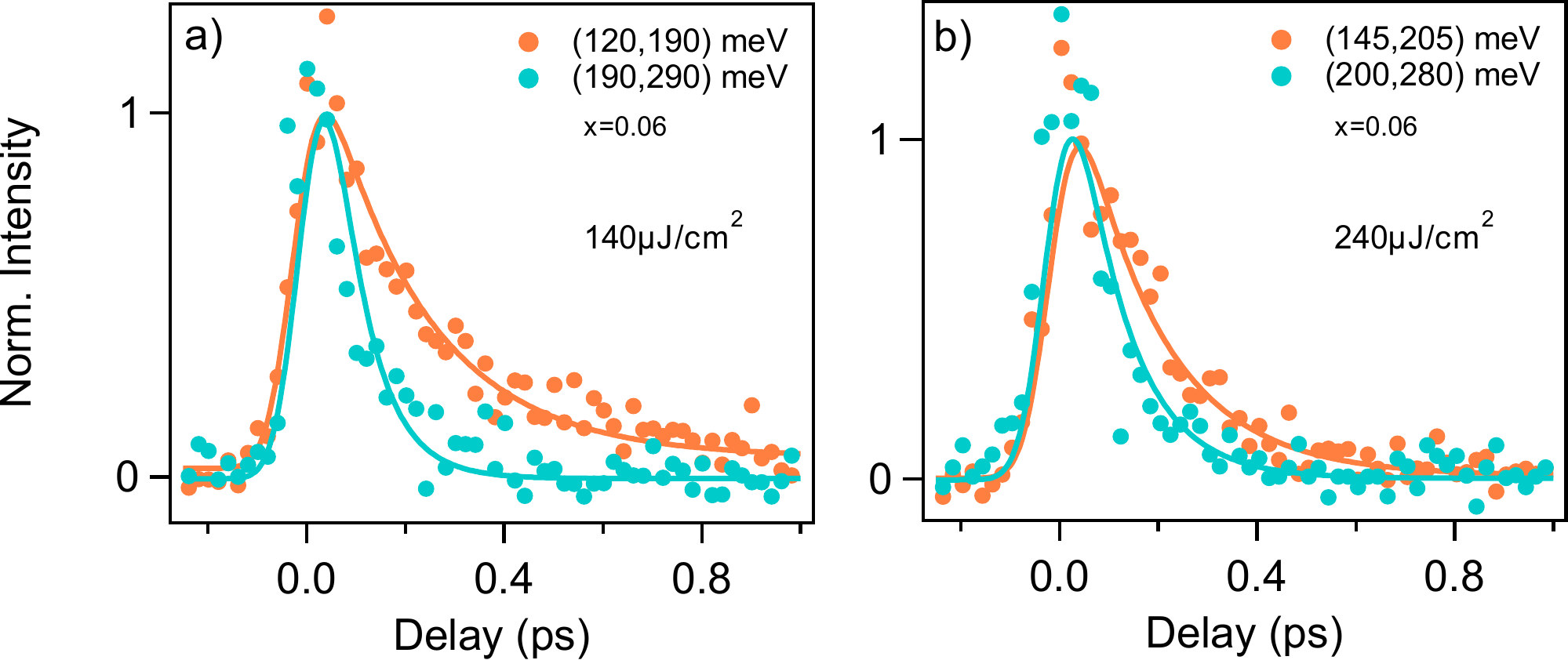}
\caption{\label{fig:f3}\footnotesize{Normalized time-dependent intensity of the $x=0.06$ (OP) compound for subsequent energy intervals around $\sim200\,\mathrm{meV}$ for the fluences of a) $140\,\mathrm{\mu J/cm^2}$ and b) $240\,\mathrm{\mu J/cm^2}$. Solid lines are fit to the data.}}
\end{figure}



\begin{figure*}[htp]
\resizebox{0.9\textwidth}{!}{\includegraphics{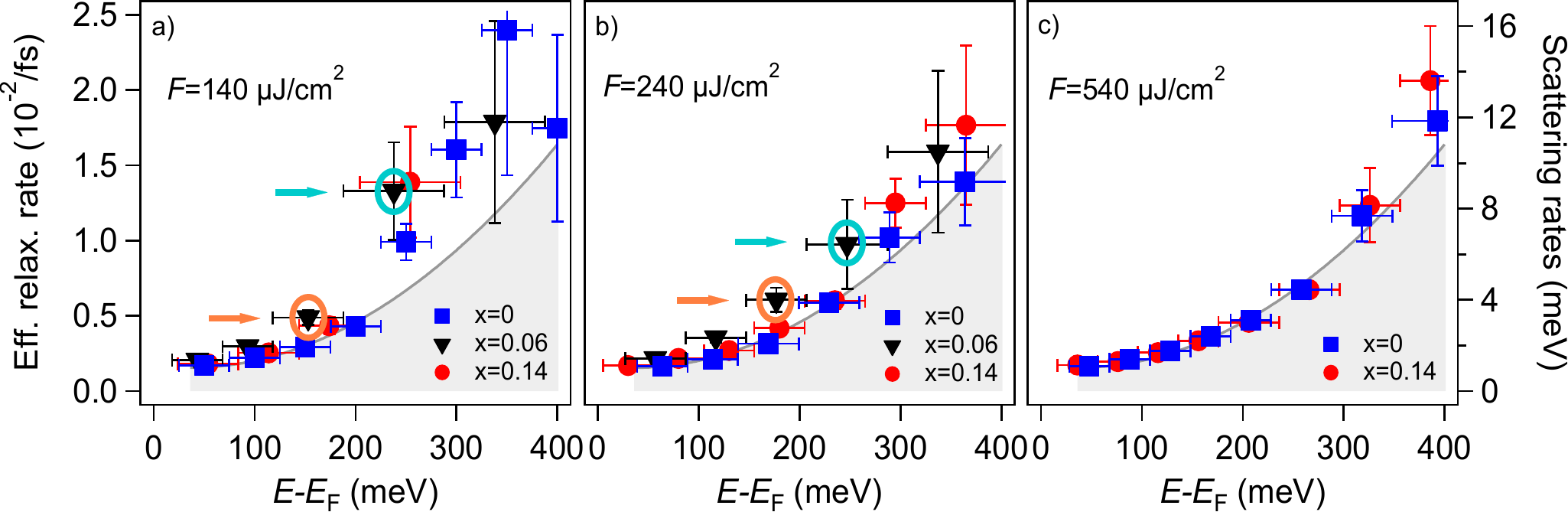}}
\caption{\label{fig:f4}\footnotesize{Relaxation rates determined by taking the inverse of the measured relaxation times vs binding energy for different fluences. The blue and orange circles highlight the data point extracted from the data in Fig. \ref{fig:f3}. Values of scattering rates in $\,\mathrm{meV}$ are also reported on the right axis in c). The contour of the grey-shaded area present in all the three graphs is a guide to the eye illustrating the high $F$ behavior, and is shown in order to emphasize the different $F$-dependence for the scattering rates at $E>200\,\mathrm{meV}$.}}
\end{figure*}

Two effects are noteworthy: the dependence of the relaxation times on $x$ does not monotonously follow the carrier concentration and the fastest relaxation is found at optimal doping. In the limit of Fermi liquid theory for a three dimensional metal, the scattering probability due to $e-e$ scattering is proportional to $n^{-5/6}$, with $n$ being the number of carriers~\cite{Quinn(1958)}. Therefore, at $T>T_N$, where the system is metallic, one would expect a monotonic dependence of the relaxation on $x$.
Interestingly similar anomalies were also found in transport measurements in the resistivity ~\cite{Katayama(2009)}. The derived scattering times~\cite{Rullier-Albenque(2009)} do not scale monotonically with Co-substitution and are shorter at optimal doping than at overdoping.\\
Another interesting point to consider is that the difference in the energy-dependent relaxation times for different Co-substitution is only present until $E-E_\text{F}=250\,\mathrm{meV}$. At this energy a change in the curves of $\tau$ vs $E-E_\text{F}$ appears as a discontinuity, which is more pronounced at the lowest $F$ as marked by the arrow in Fig. \ref{fig:fig2}(d) and absent at the maximal $F$. A closer look at data in that energy region, is instructive. In Fig. \ref{fig:f3} time dependent intensities of the $x=0.06$ compound for energy windows centered around $150\,\mathrm{meV}$ (orange) and $250\,\mathrm{meV}$ (light blue) are compared for $140\,\mathrm{mJ/cm^2}$ (Fig. \ref{fig:f3}(a)) and for $240\,\mathrm{\mu J/cm^2}$ (Fig. \ref{fig:f3}(b)). At the lowest $F$ the difference in relaxation time between the two subsequent energy windows is obvious, while it is weaker at the intermediate $F$. Such behavior becomes also evident in Fig. \ref{fig:f4} where relaxation rates $\Gamma=1/\tau$ as function of binding energy are shown. At the lowest $F$ (Fig. \ref{fig:f4}(a)) a step-like increase in the energy-dependent relaxation rates from $E-E_{\text{F}}<200\,\mathrm{meV}$ to the next values above this energy is observed for all $x$. Interestingly, the visibility of the step is reduced with increasing $F$ (Fig. \ref{fig:f4} (b)) until it completely vanishes (Fig. \ref{fig:f4} (c)).


\section{\label{discussion}Discussion}

For the investigated compounds we can reasonably assume a rigid band model~\cite{Ideta(2013)}, that is, the effect of electron doping is to add carriers to the system shifting the Fermi level upwards causing the electron pockets to enlarge and the hole pockets to reduce until they completely disappear below $E_{\text{F}}$ around $x=0.15$. In this doping region ($0<x<0.15$), where two types of competing carriers are present, other properties than for a conventional metal are found.
As a first result we have shown that at $0<E-E_\text{F}<250\,\mathrm{meV}$ the energy-dependent relaxation times depend on doping and do not scale monotonically with the carrier concentration, having the optimally doped compound the fastest decay.
It is not straightforward to unambiguously identify the relaxation channels responsible for this behavior, as for an excited population close to the Fermi level various interactions contribute to the relaxation in a complex manner. The dynamic redistribution of excited electrons has to be considered in addition to electron-boson scattering. However, we proceed by comparing our results with the ones obtained by other experimental techniques.
Static ARPES studies on the same class of compounds investigated in this work (as well as P-doped and NaFe$_{(1-x)}$Co$_x$As) report values of the imaginary part of the self-energy $\Sigma^{\shortparallel}$, obtained from linewidth ($\Gamma$) analysis for binding energy ranging from $E_\text{F}$ down to $150\,\mathrm{meV}$~\cite{Rienks(2012), Fink(2015)}. The scattering rates, obtained from $\Sigma^{\shortparallel}=\Gamma/2$, are independent of doping and increase linearly from $5\,\mathrm{meV}$ to $150\,\mathrm{meV}$. The  scattering probabilities obtained in the present work, in the time domain, from $\Gamma=\hbar/\tau$, shown on the right axis of Fig. \ref{fig:f4}, are of the order of $1\,\mathrm{meV}$ at energies $0<E-E_\text{F}<200\,\mathrm{meV}$. This difference of about two orders of magnitude is in agreement with~\cite{Yang(2015)} where the same analysis was applied to a cuprate HTSC and is a confirmation of the fact that the lifetimes measured by static ARPES and the relaxation times obtained here close to $E_\text{F}$ represent two different quantities (see also \ref{exp_res}). In the case of trARPES, relaxation of the laser-excited non-equilibrium electron population is analyzed, while in static ARPES the scattering probability of a single particle is determined. In addition, one should recall that static ARPES probes occupied states, while trARPES in the present case probes unoccupied states. Since it is known that the self energy above and below $E_\text{F}$ is different for the case of hole-doped compounds~\cite{Werner(2012)}, we can expect some differences in the respective relaxation times also in the present case.\\
On the other hand, in the doping range under investigation, qualitative agreement is found with transport measurements, where the same trend with doping found in this work is observed in resistivity measurements, and the determined $e-e$ scattering times~\cite{Katayama(2009), Rullier-Albenque(2009)} are larger near optimal doping than at over doping.\\
A widely shared interpretation~\cite{Katayama(2009),Chu(2010),Kasahara(2010),Nakajima(2011),Hashimoto(2012),Analytis(2014)}, sees antiferromagnetic fluctuations as being mainly responsible for these anomalies, because they occur near the QCP. The deviation of the $T$-dependent term of the resistivity from a quadratic to a linear scaling, accounting for the non-Fermi liquid behavior in the normal state, would qualitatively alter the $T$-dependence of the relaxation rate in the quantum critical regime, leading to the maximum scattering rate at optimal doping discussed above.
Alternative explanation for the strange normal state properties of these systems near a QCP involves the influence of a Lifshitz transition, occurring at optimal doping~\cite{Thirupathaiah(2011), Liu(2010)}, and correlation effects on the scattering rates of the charge carriers~\cite{Fink(2015), Fink(2016)}.\\
Given these considerations, the deviation from the behavior of a conventional metal which we observe in our trARPES study is not surprising, considering that FL theory does not hold. In discussing the relaxation process, the 3D character of the Fermi surface in these materials has to be taken into account. The electronic structure for these compounds is strongly doping dependent~\cite{Thirupathaiah(2010)} and, in particular, the $k_z$ dispersion increases with increasing Co concentration. That means that the observation of a Lifshitz transition, where the top of the hole pocket just touches the Fermi level is detectable only at $k_z=0$, in the case of optimal doping. The phase space for relaxation is complex and a complete discussion is beyond the scope of this paper, nevertheless we have to consider that, due to our probe pulse energy, we work at a finite $k_z$ value corresponding to a region between $\Gamma$ and $Z$ where we detect some remnant of the hole pocket in the case of the optimally doped compound, providing relaxation phase-space for $e-e$ scattering. Moreover, although we work at $T>T_N$, interband scattering (coupling between electron and hole pockets) has to be considered as a possible additional contribution to the relaxation, which is instead absent in the case of the overdoped compound, in agreement with our observation of a faster relaxation at optimal doping.
Although from our results we cannot quantify the intra- and interband conditions discussed above, we remark that despite ARPES and trARPES are closely related, qualitative agreement is found with results obtained by transport, another dynamical method.

Our second important observation comes from the analysis of the energy dependent relaxation rates at binding energies higher than those accessed by static ARPES. A step in the relaxation rates is recognized at low $F$ (Fig. \ref{fig:f4} (a)) around $E-E_{\text{F}}=200\,\mathrm{meV}$ for all $x$. This is an indication that at these energies an additional relaxation channel contribute, causing the system to suddenly relax faster. Band structure calculations predict the presence of an unoccupied $4p$ As band, at $\Gamma$, to be $0.1$ to $0.3\,\mathrm{eV}$ above the Fermi level~\cite{Fink(2009), Kordyuk(2013)}. This could provide phase space for additional inelastic interband scattering due to a peak in the density of states around that energy. However, if this was the origin of the step visible in our data, we would expect the central energy at which the step occurs to change with doping, due to the shift of the chemical potential. In addition to that, no signature of such a feature is visible in our trARPES intensities spectra (analogous to Fig. \ref{fig:fig1} (b)) for any of the dopings and fluences.
A second possibility, which we assume as the most plausible, is that an additional relaxation channel is provided by a bosonic excitation of a certain energy $\hbar\Omega$ that would couple to the excited electronic system. Theoretical modeling of the effects of electron-boson coupling ($e-b$) in trARPES~\cite{Sentef(2013)} has shown to lead to a characteristic change in the decay rates, in the form of a step. The fluence dependence of the step is a further confirmation of the presence of $e-b$ coupling, as we explain below. Electrons excited at $E-E_\text{F}>\hbar\Omega$ and the respective holes injected at $E-E_\text{F}<-\hbar\Omega$ would relax faster than those excited within the energy region $-\hbar\Omega<E-E_\text{F}<\hbar\Omega$, as shown originally from ~\cite{Engelsberg(1963)} in the case of a coupled electron-phonon system, and subsequently by Kemper et al.~\cite{Kemper(2014)} for electron-boson coupling in general. According to this description, at equilibrium, scattering from inside the bosonic energy region to states below the Fermi level is suppressed, as the electron occupation does not provide the necessary phase space for the excited electrons to relax back to energies below $E_\text{F}$. Therefore the relaxation times inside the boson window are larger compared to those at energies outside, at $E-E_\text{F}>\hbar\Omega$. In the time-domain analysis the infrared pump redistributes the electronic population, which modifies the phase space. In particular, relaxation channels which were blocked in equilibrium become now available, leading to shorter relaxation times under non-equilibrium conditions. At small $F$ one expects a situation similar to equilibrium where the signatures of the boson window in the relaxation times and the electron population are only weakly perturbed~\cite{Sentef(2013), Kemper(2014)}. Indeed our data in Fig. \ref{fig:f4}(a) show that at the smallest excitation density the step is larger showing higher (lower) scattering probability above (below) a boson energy of $\hbar\Omega=200\,\mathrm{meV}$. With increasing $F$ (Fig. \ref{fig:f4} (b)) this step reduces as the pump-induced phase space redistribution provides more relaxation channels until the signature of the boson window completely disappears at the highest $F$ (Fig. \ref{fig:f4} (c)). Such an effect was also experimentally observed in the case of cuprates~\cite{Rameau(2015)}.\\
To assign the origin of the bosonic excitation, we exclude phonons as a possible coupling candidate because the energy of $200\,\mathrm{meV}$ at which the effect is found exceeds the high energy cutoff of the phononic spectrum, around $35\,\mathrm{meV}$~\cite{Boeri(2008), Boeri(2010)}. A possible candidate is then a boson of magnetic origin. Spin fluctuations in this high energy region were reported in literature for electron doped compounds~\cite{Liu(2012), Wang(2013)} and for the parent compound~\cite{Harriger(2011)}, where they are shown to persist also at $T>T_N$. Electron-magnon coupling was also observed in ARPES measurements on ferromagnetic Fe~\cite{Schaefer(2004)}, showing a saturation of the increase of the imaginary part of the self-energy at $160\,\mathrm{meV}$, consistent with the energy of spin waves in Fe.

\section{\label{conclusion}Conclusion}

We have performed a femtosecond trARPES study on Co-substituted $\text{BaFe}_2\text{As}_2$ with different doping concentration. From the analysis of the energy dependent relaxation times we found that the optimally compound relaxes faster than the overdoped and parent compound. Our findings are therefore in qualitative agreement with the doping-dependent results of transport measurements and interband coupling between electron and hole pockets has to be considered as a possible contribution in addition to intraband relaxation, due to the finite relaxation phase-space at optimal doping, because we probe at a finite $k_z$. However, it should be noted that no $k_z$ dependence of the scattering rates was observed in static ARPES. From the analysis on the energy dependent relaxation rates we could clearly identify the presence of a step around the energy $E-E_\text{F}=200\,\mathrm{meV}$, which is best observed for small laser pump fluences. We conclude that this indicates the coupling of the excited electronic system to a bosonic excitation of magnetic origin.\\
We have demonstrated that trARPES studies are a valid contribution in elucidating the underlying mechanisms present in these complex materials, complementary to other experimental technique discussed here, although further study will be required.


\section*{Acknowledgements}

This work was funded by the Deutsche Forschungsgemeinschaft through the priority program SPP1458 and the European Union within the seventh Framework
Program under Grant No. 280555 (GO FAST). We acknowledge fruitful discussion with A. F. Kemper.


\bibliography{BaFe2As2_pdf}

\end{document}